\documentclass[twoside]{dis08}
\usepackage[latin1]{inputenc}
\usepackage[dvips]{graphicx,epsfig,color}
\usepackage{wrapfig,rotating}
\usepackage{amssymb,amsmath,array}

\begin{document}
\title{Inclusive diffraction with the ZEUS detector at HERA: 
a comparison among selection methods}

\author{Marta Ruspa
%
%
\vspace{.3cm}\\
%
Univ. Piemonte Orientale and INFN-Torino, Italy
%
\vspace{.1cm}\\
}

\newcommand{\spom} {\mbox{$\scriptstyle \mathrm{I}\! \mathrm{P}$}}
\maketitle

\begin{abstract}
The diffractive 
dissociation of virtual photons, $\gamma^{\star}p \to Xp$, has been studied 
with the ZEUS detector at HERA by requiring a large rapidity gap between 
$X$ and the outgoing proton, by analysing the mass 
distribution, $M_X$, of the hadronic final state, as well as by directly 
tagging the proton. At low values of the proton momentum loss, 
the diffractive structure function measurements obtained with the three 
methods are consistent, provided the different treatment and contributions 
of proton-dissociative events are taken into account.
\end{abstract}

\section{Inclusive diffraction at HERA}

In diffractive interactions in hadron-hadron or photon-hadron 
collisions at least one of the beam particles emerges intact from 
the collision, having lost only a small fraction of its initial energy, 
and carrying a small transverse momentum. 
Such interactions are described by the exchange of an object with 
vacuum quantum numbers, referred to 
as the Pomeron in the framework of Regge phenomenology~\cite{regge}. 
Similar reactions can also proceed when quantum 
numbers are exchanged through subleading Reggeon and pion trajectories; 
however, these contributions 
are negligible at small values of the energy loss. 

Significant progress has been made in understanding diffraction 
in terms of Quantum Chromo-Dynamics by studying 
the diffractive 
dissociation of virtual photons, $\gamma^{\star}p \to Xp$, in deep 
inelastic $ep$ scattering (DIS) at HERA (for a review see~\cite{review}). 
In this process, a photon of virtuality $Q^2$ 
diffractively dissociates interacting with the proton at a 
centre-of-mass energy $W$ 
and produces the hadronic system $X$ with mass $M_X$. 
The fraction of the proton's momentum carried by the exchanged 
object is denoted by $x_{\spom}$, while the fraction of the momentum
 of the exchanged object carried by the struck quark is denoted by 
$\beta$. 

\section{Comparison between selection methods}

Experimentally, diffractive $ep$ scattering 
is characterised by the presence of a leading proton in the 
final state carrying most of the proton beam energy and, 
consequently, by a lack of hadronic activity in the 
forward (proton) direction. Conservation of momentum implies that 
the system $X$ must have a small mass with respect to the 
photon-proton centre-of-mass energy, since $x_{\spom}~\gtrsim~M_X^2/W^2$. 
These signatures have been widely exploited at HERA to select 
diffractive events by tagging the foward proton 
(proton-tagging method~\cite{ptagging}), 
by requiring the presence of a large gap in the forward rapidity 
distribution of particles (LRG method~\cite{LRG}) or 
by exploting the shape of the $M_X$ distribution, different 
in diffractive and non-diffractive events 
($M_X$ method~\cite{mx,mx1}). 

A thorough comparison of these three selection methods has recently 
been carried out on a set of data collected with the ZEUS detector  
in the years 1999 and 2000~\cite{mx2,lrgnoi}, when the detector 
was still equipped with the leading proton spectrometer (LPS).  
The different methods access different kinematic regions and are subject 
to different systematic uncertainties: in the LRG and $M_X$ methods, 
high $M_X$ values are not accessible since the non-diffractive 
background grows with $M_X$ and the rapidity gap moves more 
and more forward (and is eventually confined to the beam pipe). 
Moreover, the measured cross section includes a contribution from 
events of the type $ep \rightarrow eXN$, in which the proton also 
dissociates into a state $N$ with low-mass $M_N$, 
separated from $X$ by a rapidity gap. 
The statistical precision of the results is good because of 
the high acceptance of the central detector. 
Conversely, low-$x_{\spom}$ samples selected by the proton-tagging 
method have little or no background from proton-dissociative events or from 
non-diffractive DIS and allow access to higher values of $M_X$.  
However, the statistical precision is poor 
because of the small acceptance of the proton taggers 
-- approximately 2\% at low $x_{\spom}$ in the LPS case.

In order to compare the reduced cross sections measured with the 
three methods, the different $x_{\spom}$ and $M_N$ coverages of 
the various samples have to be taken into account: 

\begin{itemize}

\vspace{-0.22cm}
\item the LPS data extend up to $x_{\spom}$ of 0.1 and therefore 
include contributions from Reggeon and pion trajectories; the 
LRG sample is restricted to the region $x_{\spom} < 0.02$ and 
thus mainly consists of diffractive events; in the $M_X$ sample, 
the statistical subtraction of the non-diffractive events has been 
shown~\cite{mx1} to suppress the Reggeon contribution;

\vspace{-0.15cm}
\item in the LPS results, $M_N$ coincides with the protons 
mass, $M_p$; the LRG data are also corrected to $M_N = M_p$; 
the $M_X$ results are corrected to $M_N < 2.3$ GeV. 

\vspace{-0.2cm}
\end{itemize}

\vspace{-0.05cm}
The amount of proton-dissociation background and the corresponding 
corrections were found to be the most crucial issue in the 
comparison and are therefore discussed in detail in the following.  

\subsection{Proton-dissociative background and relative corrections}
\label{sec:pdiss}

The proton-dissociative system can either escape 
entirely undetected in the forward beam-pipe or leak partially 
into the detector acceptance and therefore be measured by the 
forward detectors (forward plug calorimeter, FPC, and main calorimeter, 
CAL). In the former case, the background events are included in the 
measured cross section, of which they bias the normalisation. 
In the latter case, they are or are not rejected depending on the 
specific analysis cuts. 

In the LPS analysis 
the contribution from proton-dissociative events 
was studied with the {\sc Pythia} Monte Carlo (MC) 
and was found around 9\% at $x_{\spom}=0.1$, 
decreasing rapidly with decreasing $x_{\spom}$. In the region 
$x_{\spom} <0.02$ this background is negligible. 
At low $x_{\spom}$ the ratio of the LRG and $M_X$ results 
to the LPS ones can thus be used to quantify the total fraction 
of proton-dissociative events included in these samples.

In the LRG analysis 
the contribution from proton-dissociative events was also estimated with 
{\sc Pythia}. Two proton-dissociative samples were selected, 
one with the FPC and one with the LPS, the combination of which 
covers nearly the whole $M_N$ spectrum, including the lowest $M_N$ values. 
The generated {\sc Pythia} distributions for $M_N$, $M_X$ and $Q^2$ were 
reweighted to describe these samples, in particular 
the energy distribution in the FPC and the 
$x_L$ distribution in the LPS. 
The average of the FPC and LPS estimates provided a measurement of the 
proton-dissociative contribution to the LRG results of $25 \pm 1 {\rm 
(stat.)} \pm 3 {\rm (syst.)}\%$.

In the $M_X$ method the diffractive contribution is extracted as 
the observed number of events after subtracting the non-diffractive 
component from a fit to the $lnM_X^2$ distribution. 
Proton-dissociative events measurable 
in the forward detectors lead to a reconstructed $M_X$
value higher than the actual photon-dissociative mass, hence 
to a distortion of the $lnM_X^2$ spectrum and, consequently, of the 
extracted diffractive signal. The role and the treatment of the 
proton-dissociation background is thus much more critical than in 
the LRG analysis: before the statistical subtraction of the 
non-diffractive background, all events need to be subtracted 
from the data which, according to a proton-dissociative MC, deposit 
anything measurable in the forward detectors. It has been 
shown~\cite{heuijin_thesis} that, on average, events with masses 
$M_N < 2.3$ GeV cannot be detected. 
The {\sc Sang} MC was used to subtract 
events with $M_N > 2.3$ GeV from the data. To avoid the Reggeon exchange 
region {\sc Sang}, like most of the proton-dissociative MCs, is generated 
with an upper $M_N$ cut, $(M_N/W)^2 < 0.1$. Hence, bin-by-bin the amount of 
subtracted events with $M_N > 2.3$ GeV depended on $W$. 
On the contrary, Monte Carlo studies showed that in the LRG case the 
rapidity gap requirement eliminates the $M_N$ tail; the 
correction becomes therefore independent of kinematics, as 
discussed in next Section and shown in Fig.~\ref{fig:ratio}.

\section{Comparison between cross section results}
\label{sec-comp}

\begin{wrapfigure}{r}{0.45\columnwidth}
\vspace{-1.2cm}
\centerline{\includegraphics[width=0.33\textheight]{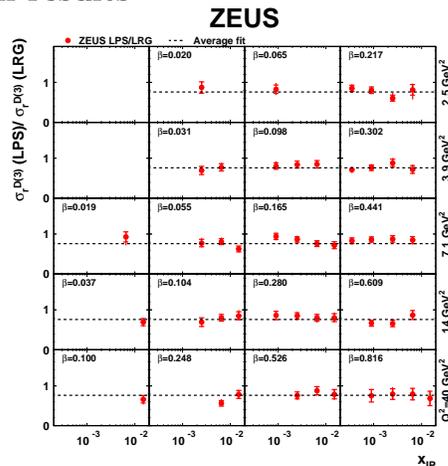}}
\vspace{-0.5cm}
\caption{Ratio LPS/LRG.}
\label{fig:ratio}
\end{wrapfigure}
The results obtained with the three methods were compared in bins of 
$M_X$, $Q^2$ and $x_{\spom}$ in terms of the diffractive reduced 
cross section, $\sigma_r^{D(3)}$. The latter coincides with the diffractive 
structure function, $F_2^{D(3)}$, if the ratio of the cross sections 
for longitudinally and transversely polarised virtual photons can be 
neglected.  
The three samples are only weakly correlated through systematics 
but statistically not independent: 
the LRG and $M_X$ data overlap by about 75\%; 
0.7\% of the LRG events have a proton measured in the LPS 
and 35\% of the LPS events are also contained in the LRG sample.

The ratio of $\sigma_r^{D(3)}$, extracted from the LPS and LRG data, 
shown in Fig.~\ref{fig:ratio},  
is $0.76 \pm 0.01 {\rm (stat.)} ^{+0.03} _{-0.02} 
{\rm (syst.)} ^{+0.08} _{-0.05} {\rm (norm.)} $; 
the last uncertainty reflects the normalisation uncertainty of the 
LPS data, mostly related to the $\pm 7\%$ uncertainty due to the 
proton-beam optics. 
The ratio is independent of 
$Q^2$, $x_{\spom}$ and $\beta$, indicating that the two methods lead 
to compatible results for $x_{\spom} < 0.01$. 
It also confirms that contributions from 
proton-dissociative events in the LRG measurement do not significantly 
alter the $Q^2$, $x_{\spom}$ or $\beta$ 
dependences. The ratio translates into a proton-dissociative 
background fraction of $24 \pm 1 
{\rm (stat.)} ^{+2} _{-3} {\rm (syst.)} ^{+5} _{-8}{\rm (norm.)} $\%. 
The agreement between this number and the result of the MC study discussed 
in Sec.~\ref{sec:pdiss} lends support to the present estimate of the proton-
dissociation contamination in the LRG analysis.

\begin{figure}[hb]
  \includegraphics[height=.41\textheight]{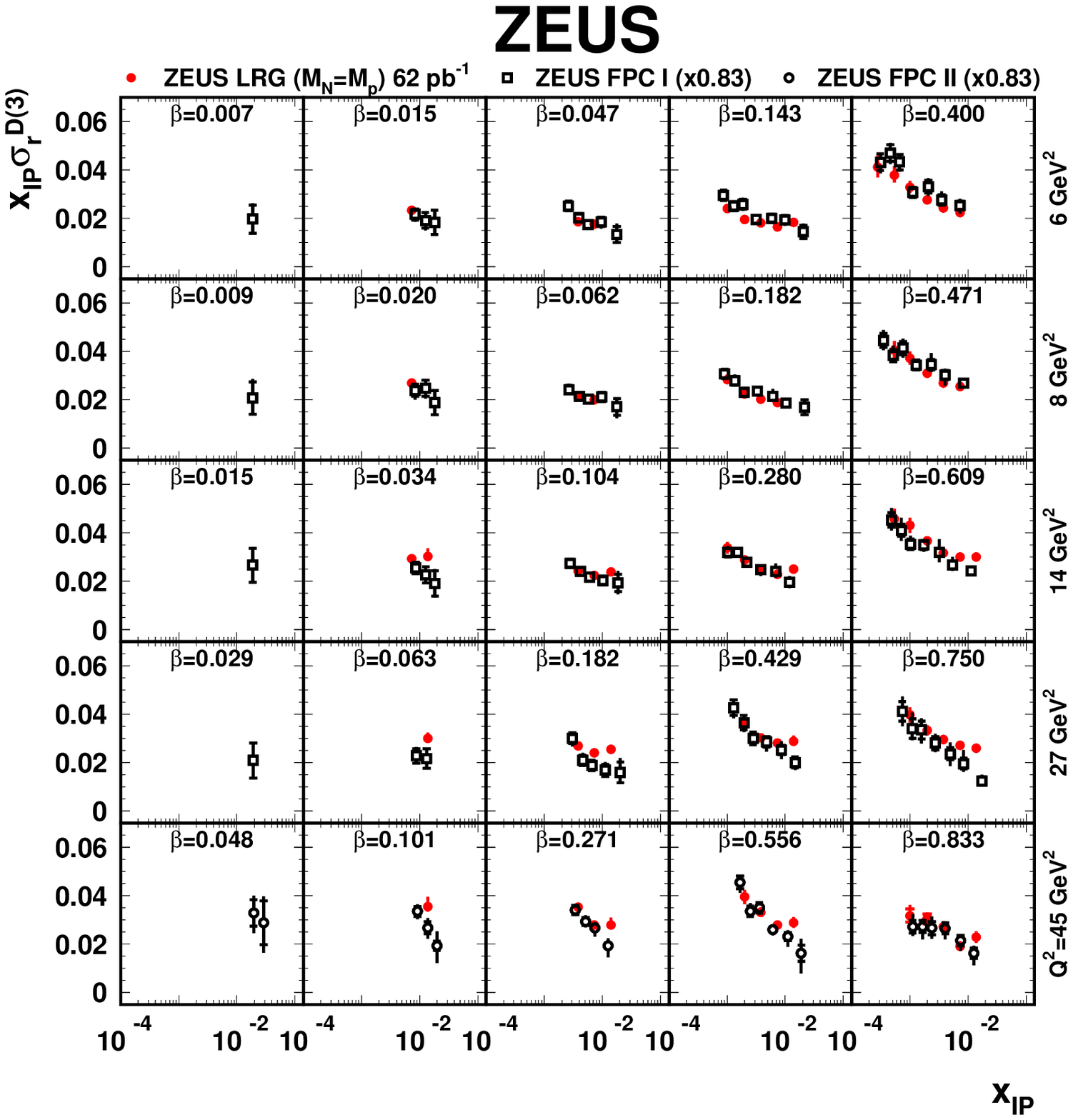}
  \hspace{-0.6cm}
  \includegraphics[height=.41\textheight]{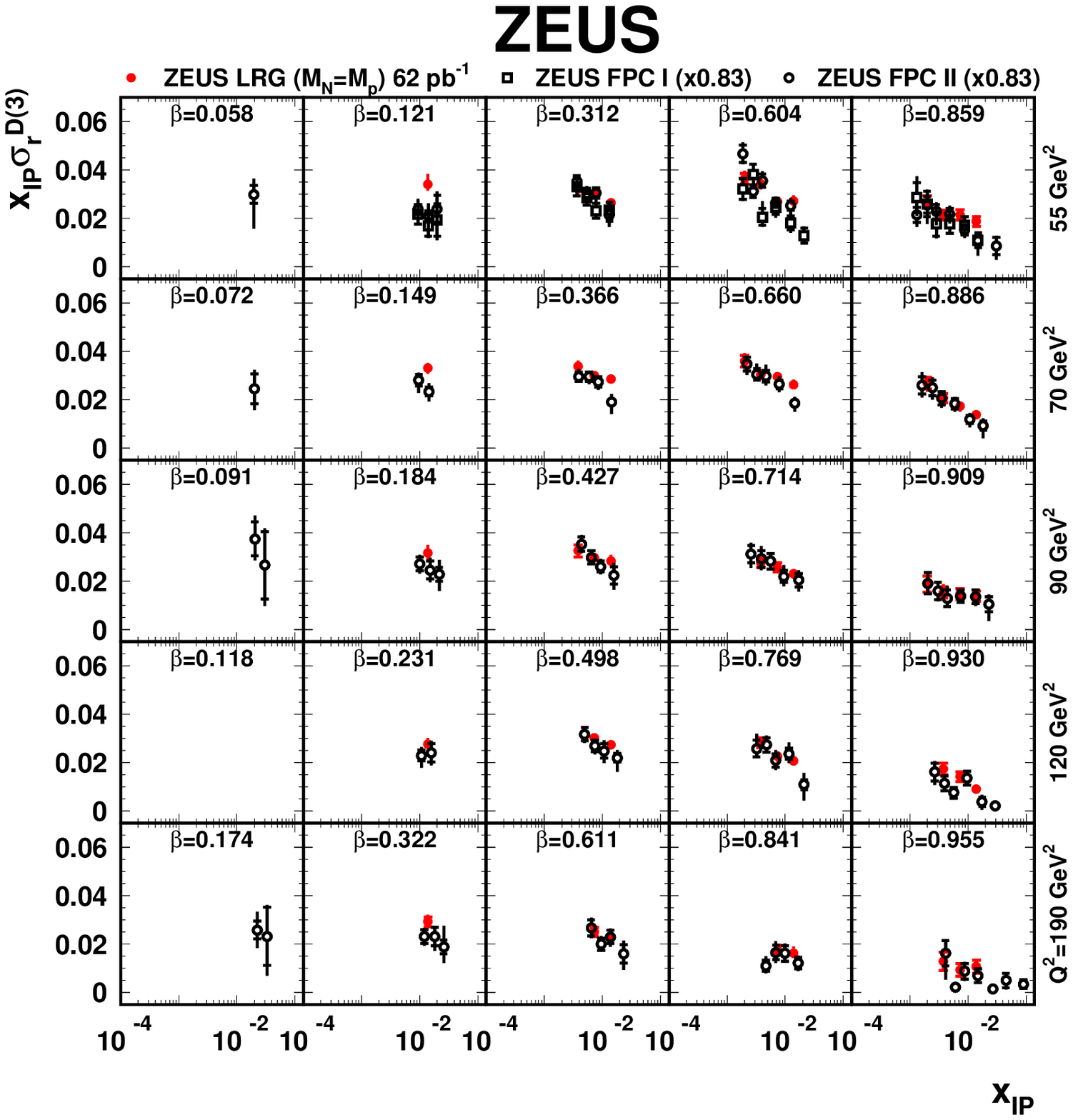}
\vspace{-0.8cm}
  \caption{Comparison $M_X$-LRG.}
  \label{fig:comp_mx}
\end{figure}

Cross section measurements 
obtained with the LRG and $M_X$ methods are compared  
 in Fig.~\ref{fig:comp_mx}, where also 
the previous $M_X$-method results~\cite{mx1} are shown. The LRG data  
were corrected to $M_N=M_p$ by statistical subtraction of the background 
estimated in Sec.~\ref{sec:pdiss}. The $M_X$ 
results were normalised to the LRG data with a scaling factor of 
0.83 $\pm$ 0.04, obtained from a global fit; this factor 
quantifies the amount of residual proton-dissociative background in 
the $M_X$ data due to masses below $2.3$ GeV. The overall agreement 
between the two measurements 
is satisfactory. The different $x_{\spom}$ dependence for $x_{\spom}\gtrsim0.01$, more evident at low 
$Q^2$, may be ascribed to the fact that in the $M_X$ results the 
contribution of Reggeon and pion trajectories is suppressed. 
In the low-$Q^2$ 
region, the $Q^2$ behaviour is slightly different in the two data 
sets, with the $M_X$-method results decreasing faster with $Q^2$ than the 
LRG results.

\vspace{-0.4cm}

\vspace{-0.5cm}



\end{document}